\documentclass[preprint,12pt,authoryear]{elsarticle}

\usepackage{amssymb}
\usepackage{amsmath}
\usepackage{booktabs}

\journal{   }

\begin{document}

\begin{frontmatter}



\title{Prediction of Daily $PM_{2.5}$ Concentration in China Using Data-Driven Ordinary Differential Equations}

 \author[label1]{Yufang Wang}
 \author[label2]{Haiyan Wang}
 \author[label3]{Shuhua Zhang}
 \address[label1]{Department of Statistics,
        Tianjin University of Finance and Economics, Tianjin 300222, China}
 \address[label2]{Mathematical and Natural Sciences,
     Arizona State University, AZ 85069,  USA}
 \address[label3]{Coordinated Innovation Center for
        Computable Modeling in Management Science,
        Tianjin University of Finance and Economics, Tianjin 300222, China}

\begin{abstract}
Accurate reporting and forecasting of $PM_{2.5}$ concentration are important for improving public health. In this paper, we propose a  daily prediction method of $PM_{2.5}$  concentration by using data-driven ordinary differential equation (ODE) models. Specifically, based on the historical $PM_{2.5}$  concentration, this method combines genetic programming and orthogonal least square method to evolve the ODE models, which describe the transport of $PM_{2.5}$ and then uses the data-driven ODEs to predict the air quality in the future.  Experiment results show that the ODE models obtain similar prediction results as the typical statistical model, and the prediction results from this method are relatively good. To our knowledge, this is the first attempt to evolve data-driven ODE  models to study $PM_{2.5}$ prediction.
\end{abstract}

\begin{keyword}
concentration data \sep genetic programming \sep
least square method \sep ODE models \sep $PM_{2.5}$ prediction
\end{keyword}

\end{frontmatter}


\section{Introduction}
Air pollution has became one of the most challenge environmental problems.
$PM_{2.5}$ (particulate matter smaller than 2.5 $\mu$m) has been found to play a significant role for decreasing visibility, negative effects on human health, and influence on air pollution.  Accurate and timely forecasting of PM2.5 concentration is essential for improving public health and economic conditions.
Because $PM_{2.5}$ concentrations are dynamic and exhibit wide variation for different cities in China, accurate description and prediction in China become a highly challenging task for scientists.

There is a host of studies on $PM_{2.5}$ prediction.  Each approach addresses the problem from different perspectives. Physico-chemical methods and satellite remote sensing techniques are widely used in  Meteorological science. For example, 3-D chemistry transport models (CTMs) mainly address the formation mechanism of $PM_{2.5}$ from the view of physico-chemical and meteorological processes through the temporal dynamics of the emission quantities of various pollutants (\cite{yahya2014real,chuang2011application}). This approach needs a large number of various meteorological data and perfect representation of the physico-chemical processes; therefore, it is difficult to guarantee real-time forecast. Satellite remote sensing techniques  have the advantages of spatially seamless and long-term coverage; as s result, in recent years they have been widely employed to predict $PM_{2.5}$ by considering satellite-derived aerosol optical depth empirically correlated with $PM_{2.5}$  (\cite{ma2014estimating}). However, the equipment expense for this type of research is relatively high.

Statistical approach is a very popular empirical prediction method. It aims to detect  certain correlated patterns between air quality data and various selected predictors, thereby predicting the pollutant concentrations in future. Common statistical approaches, such as linear regression models (\cite{li2011study,benas2013estimation}), neural networks (\cite{mao2017prediction}), nonlinear regression models (\cite{emili2010pm10}) and  neurofuzzy models, are easier to implement but limited to specific geographical locations.
In our previous work (\cite{wang2018prediction}), a partial differential equation (PDE) model, specifically, a linear diffusive equation, was applied to describe the spatial-temporal characteristics of $PM_{2.5}$ for short-term prediction.
Average prediction accuracy of the PDE model over all city-regions is $93\%$ or $83\%$ with different accuracy definitions. We  use a simple logistic growth PDE model with reasonable assumptions based on meteorological knowledge and applied mathematics knowledge.
In this paper, we present a data-driven method to improve the simple PDE for predicting $PM_{2.5}$ in China.

A large amount of available data has sprung up
in our lives. Nowadays, monitoring stations in a city can provide real-time air quality.
Evolutionary  modeling method, as a data-driven identification algorithm, is used to help build ordinary differential equation (ODE) models for $PM_{2.5}$ prediction in this paper. Genetic programming (GP)is an important EM method, which mimics the mechanisms of natural selection and genetic variation.
Based on some suitable coding, GP uses genetic operators and the principle of ``survial of the fittest'' to search for the optimal solutions. An evolutionary modeling method of ODEs with GP is proposed in \cite{cao2000evolutionary} and \cite{chen2011time}, in which a genetic algorithm is applied to optimize the parameters of a model. Compared with genetic algorithm for parameter optimization, least square method  can analytically calculate the linear-in-parameter models. Therefore, in \cite{madar2005genetic}
  a GP method with least square method for identification of linear-in-parameter models is proposed.

In this paper, we extend the work of \cite{cao2000evolutionary} and \cite{madar2005genetic} to develop an algorithm for constructing ODEs which combines GP algorithm and orthogonal least square (OLS) algorithm. Specifically,
the ODE model will involve the concentrations $y$  of $PM_{2.5}$  varying with time $t$ and its change rate $y'(t)$  and $y''(t)$ which are related to the current concentration $y$ and the current time $t$. Therefore, the dynamic process of $PM_{2.5}$ concentration is naturally described by an ordinary differential equation,
\begin{equation}
\label{eq1}
\frac{dy}{dt}=f_1(y,t)
\end{equation}
 or
\begin{equation}
\label{eq2}
\frac{d^2y}{dt^2}=f_2(y,\frac{dy}{dt},t),
\end{equation}
but the exact mathematical formulas of $f_1(y,t)$ and $f_2(y,\frac{dy}{dt},t)$ will be determined by $PM_{2.5}$ data.
We may make some reasonable  preassumptions about the model structure. But it is almost impossible to develop a model to include all factors that affect the $PM_{2.5}$, which needs more knowledge of the specific atmosphere details. The aim of this paper is to apply genetic algorithm to identify the model structure from real data of $PM_{2.5}$ concentrations, thus to further make prediction for  $PM_{2.5}$ in the future.

The main contribution of this paper are two-fold:

\textbf{$\bullet$} We propose a novel data-driven ODE construction method based on tree-based genetic programming and OLS to predict the future  $PM_{2.5}$ concentration using historical $PM_{2.5}$ concentration observations. This method requires only short-term $PM_{2.5}$ concentration data. In addition, this method needs almost no meteorological
assumptions; therefore, it can be easily applied to other problems.

\textbf{$\bullet$} We evaluate our approach with Wuhan's $PM_{2.5}$ concentration data of about half a year from the view of in-sample and out-of-sample predictions. Compared with traditional statistical regression model, our models obtain
relative good prediction results. These experiments suggest that our new method of predicting $PM_{2.5}$ is promising.

The paper is organized as follows. Section 2 gives a general description for the ODE construction algorithm: genetic programming with the least square method.
Section 3 describes the details of the ODE construction, section 4 validates the effectiveness of the proposed model, and finally conclusions are drawn in section 5. In particular, in section 4, we will make
in-sample and out-of-sample predictions to measure the ODE models and compare the evolutionary ODE model proposed in this paper with the statistical model to demonstrate the feasibility of the ODE model.

\section{Genetic programming for ODE}

In this paper, we develop a genetic programming  algorithm to construct ODEs for  $PM_{2.5}$ prediction.
The higher-order ODE (\ref{eq2}) can be converted into an ODE system with the form of
\begin{equation}
\left\{
\begin{array}{l}
\frac{dy_1}{dt}=y_2,\\
\frac{dy_2}{dt}=f_2(y_1,y_2,t)
\end{array}
\right.
\end{equation}
If we know the construction of $f_2(y_1,y_2,t)$, we just replace $y_1$ and $y_2$ by $y$ and $\frac{dy}{dt}$ respectively; then we can easily obtain the structure of (\ref{eq2}). Therefore, the construction of  (\ref{eq2}) equals the construction of the following:
\begin{equation}
\label{eq3}
\frac{dy_2}{dt}=f_2(y_1,y_2,t),
\end{equation}
where $y_1$ satisfies $\frac{dy_1}{dt}=y_2$.
As a result, the construction of (\ref{eq1}) and (\ref{eq2}) is essentially the same problem as the construction of  one-order ODEs (\ref{eq1}) and (\ref{eq3}). In the following, we discuss only problem (\ref{eq1}). Problem (\ref{eq2}) can be discussed in the same way.

The ODE construction of (\ref{eq1}) contains the structure construction of the function $f(y,t;p)$ and parameter identification of vector $p$ from
\begin{equation}
\left\{
\begin{array}{l}
\frac{dy}{dt}=f(y,t;p),\\
y(t_0)=y_0\in\Re^1,
\end{array}
\right.
\label{inverse_problem}
\end{equation}
using additional measurements of the following type:
\begin{equation}
\label{inverse_initial}
y(t_k)=Y_k, k=0,1,2,\dots,n.
\end{equation}
We reduce problems (\ref{inverse_problem}) and (\ref{inverse_initial}) to an optimization problem, which consists in minimizing of the functional
\begin{align}
\label{Jfp}
J(f,p)=&\frac{1}{n}\Bigg\{\sum_{k=2}^{n-1}\Big(f(Y_k,t_k;p)-\frac{Y_{k+1}-Y_{k-1}}{2\triangle t} \Big)^2\notag\\
&+\Big(f(Y_1,t_1;p)-\frac{-Y_{3}+4Y_{2}-3Y_{1}}{2\triangle t}\Big)^2\notag\\
&+\Big(f(Y_n,t_n;p)-\frac{3Y_{n}-4Y_{n-1}+Y_{n-2}}{2\triangle t}\Big)^2\Bigg\},
\end{align}
where $\triangle t$ is the time interval.

The construction of the ODEs contains structure selection and parameter selection. GP is an evolutionary computation technique, which transforms the structure selection problem to a symbolic optimization problem, in which the search space consists of possible compositions of predefined symbols from the symbol set. Specifically, the construction of the ODE models is concluded as follows:

\textbf{1) Defining initial function set and operator set.}

Denote the function set as $I_1$, containing the predefined elementary functions in $f$; Denote the operator set as $I_0$, including the basic arithmetic operations existing between the elementary functions in f.

\textbf{2) Generating an initial population}

Each ODE model can be uniquely represented by a tree (\cite{cao2000evolutionary,chen2011time}).
Under the condition that the maximum tree depth does not exceed predefined constant $D$, based on the function set $I_1$ and
operator set $I_0$,
the algorithm randomly generates a lot of potential structures of $f$ in the form of tree-structure. Every $f$ is regarded as an individual of the population in GP. This is the first generation of the genetic system and the optimal ODE structure is evolved from the first generation.

\textbf{3) Structure selection.}

 We define the fitness function as (\ref{Jfp}) and it measures which of the current ODE structure is better suited to the $PM_{2.5}$ concentration.
Calculate the fitness value of every tree in the current generation, and operate mutation and crossover on the ODE-trees with lower fitness in the current generation (They are parents of the next generation). Measure the fitness of the newly-generated offsprings. Select predefined number of individuals
from all the parents and offsprings by the rule of higher fitness value, which has most wins to form the next generation.

Specifically, mark $T^{a}$ and $T^{b}$  are two ODE-trees of the current generation:

\textbf{$\bullet$ Crossing.} As the predefined crossover rate, perform crossover on the trees with the lower fitness.  Tree-level crossover performs the following operations on parent $T^{a}$ and $T^{b}$. Randomly select a node in each tree as crossover point, exchange the subtree  rooted at the crossover points and generate two new ODE-trees $T^{c}$ and $T^{d}$.

\textbf{$\bullet$ Mutating.}
According to a predetermined mutation rate, perform mutation on the trees with lower fitness. For example parent $T^{a}$, randomly select a node within the tree as the mutation point with a randomly generated tree, thus an offspring $T^{e}$ is generated.

\textbf{$\bullet$ Selecting.}
Compute the fitness value of all the parents and the newly-produced offsprings and delete the trees who have lower fitness as the number of the new generation we predefined.

\textbf{4) Parameter identification.}

At some interval of the generations, select the better structures to optimize parameter by OLS methods.

 \textbf{5) Forming new generations recursively}

Combining Step 3 and Step 4, the algorithm forms the new generation.

\textbf{6) Checking the exit conditions}

Step 3 and Step 4 are repeated in each generation until a predefined number of generations has reached or the best ODE structure is found.

More specific ODE construction, for the prediction of $PM_{2.5}$, will be described in the next section.
\section{Construction of ODEs for prediction of $PM_{2.5}$ concentration}

As described in (\ref{inverse_problem}) above, $PM_{2.5}$ concentrations can be described by a dynamical system. As the right part of the ODE model, $f(y,t;p)$ should consist of multiple elementary functions. In this section, we develop an ODE-construction algorithm by combining the genetic algorithm and the OLS method. Our goal is to construct $f(y,t;p)$ by identifying the elementary functions in $f(y,t;p)$ and associated parameters $p$.
\subsection{Genetic programming for ODE structure}
We will explain the genetic programming for constructing ODE in this subsection. For convenience, in this subsection we use specific sets of elementary functions and operations, but it can easily be expand to other sets of elementary functions and operations. Suppose that $f(y,t;p)$ can be described by four correlated functions $y(t),t,sin(t),e^{t}$ and the basic arithmetic operations between these functions are ``plus", ``minus" and ``multiply"; therefore we
denote $I_0=\{+, \times\}$ and $I_1=\{y(t),t,sin(t),e^{t}  \}$. The reason we choose these two sets is that the varying rate of $PM_{2.5}$ concentration is related to the existing concentration $y(t)$ and the time $t$. And it behaves periodically (therefore we select $sin(t)$) or shows rapid growth (therefore we select $e^{t}$) in certain weather condition.
Now suppose that a series of observed values of $y(t_i)$ are collected at the time $t_i=t_0+i\times\Delta t$, $(i=1,2,\dots,n)$; thus $X(t)=[y(t),t,sin(t),e^{t}]$ can be written as
\begin{equation}
X=
\left(
\begin{array}{cccc}
y(t_1)&t_1&sin(t_1)&e^{t_1}\\
y(t_2)&t_2&sin(t_2)&e^{t_2}\\
\vdots&\vdots&\vdots&\vdots\\
y(t_n)&t_n&sin(t_n)&e^{t_n}\\
\end{array}
\right)
\end{equation}

$dY(t)=\frac{dy}{dt}$ at time $t_i, i=1,2,\dots,n$ can be approximated by its second-order difference format as
\begin{eqnarray}
dY(t_i)=\frac{dy}{dt}(t_i)=
\begin{cases}
\frac{-Y_{i+2}+4Y_{i+1}-3Y_i}{2\triangle t}, &i=1\cr
\frac{Y_{i+1}-Y_{i-1}}{2\triangle t}, &i=2,3,\dots,n-1 \cr
\frac{3Y_{i}-4Y_{i-1}+Y_{i-2}}{2\triangle t}, &i=n
\end{cases}
\end{eqnarray}
thus $dY(t)=\frac{dy}{dt}$ can be expressed as
\begin{equation}
\label{Y}
dY=
\left(
\begin{array}{c}
\frac{dy}{dt}(t_1)\\
\frac{dy}{dt}(t_2)\\
\vdots\\
\frac{dy}{dt}(t_n)\\
\end{array}
\right)
\end{equation}

Denote $f(X)=[f(X(1,:)),f(X(2,:)),\dots,f(X(n,:))]^{T}$, where $f(X(i,:))=f(y(t_i),t_i,sin(t_i),e^{t_i})$ is the composite function of the elementary functions involving variables $y(t),t,sin(t),e^{t}$ and the function space defined by those functions can be denoted by $F$. Then the optimal problem is to find the model, having the form of
$$dY^{*}=f(X^*)$$
such that
\begin{equation}
\label{minfunction}
min\{||dY^*-dY||, \forall f \in F\},
\end{equation}
 where
$$||dY^*-dY||=\frac{1}{n}\sum_{i=1}^{n}\Big(dY(t_i)-f\big(X(i,:)\big)\Big)^2.$$

  \begin{figure}[!htbp]
\centering
 \includegraphics[width=0.4\textwidth]{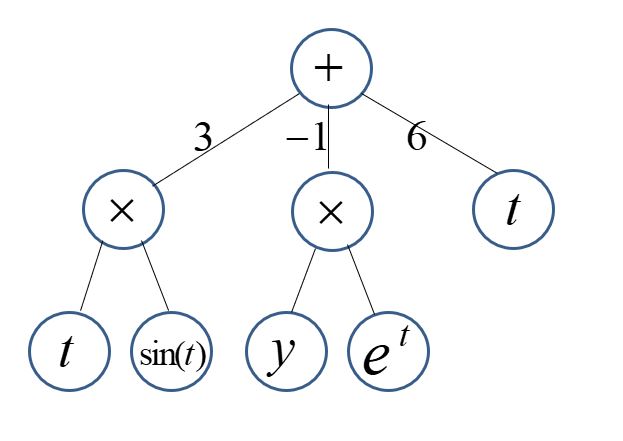}
\caption{The representation of a ODE model.}
\label{fig6}
\end{figure}

 With a predefined function set F,  for any function $f\in F$,
  it is easy to see that an ODE can be uniquely represented by a tree (\cite{cao2000evolutionary,chen2011time}).
As is the case with  $f(t, sin(t), e^t, y)$, an ODE model with the form of
\begin{equation}
\label{ft}
f(t, sin(t), e^t, y)=3tsint-ye^t+6t,
\end{equation}
can be uniquely represented by a tree as
 Figure \ref{fig6}.

Therefore, when we perform crossover, mutation and selection on the ODE-tree-model, the ODE structure $f$ in (\ref{inverse_problem})
will update till satisfying (\ref{minfunction}). And if the genetic programming is confined around a local minimum, the mutation step will help to get out of it. As practice shows, a global minimum of (\ref{minfunction}) should be obtained.

\subsection{Fitness Function}
To construct the structure of the ODE model, we discuss $f$ having the form of
\begin{equation}
\label{eq4}
f(t_k)=\sum_{i=1}^{M}p_iF_i(X(k,:)), k=1,2,\dots,n,
\end{equation}
where $F_1, F_2, \dots, F_M$ contain all the nonlinear parts of function $f$, and the parts are composed of  the values of the predefined function set at $t_k$. As is the case with (\ref{ft}), they are composed of
$t_k, sint_k, e^{t_k}, y(t_k)$ based on the operation `` minus" and `` plus". Equation (\ref{eq4}) is essentially a linear-in-parameters model. We make this assumption because an overly complex model is not conducive to describing the nature of the problem and we have put all the potential nonlinear form in the function set $I_1$, which is included in $F_i, i=1,2,\dots,M.$

In this paper, we define the fitness function as
\begin{align*}
Fitness
=&\frac{1}{n}\sum_{k=1}^{n}\Big(\frac{dy}{dt}(t_k)-\sum_{i=1}^{M}p_iF_i\Big(X(k,:)\Big)\Big)^2\\
=&\frac{1}{n}\Bigg\{\sum_{k=2}^{n-1}\Big(\frac{Y_{k+1}-Y_{k-1}}{2\triangle t} -\sum_{i=1}^{M}p_iF_i\big(X(k,:)\big)\Big)^2\\
&+\Big(\frac{-Y_{3}+4Y_{2}-3Y_{1}}{2\triangle t}-\sum_{i=1}^{M}p_iF_i\big(X(1,:)\big)\Big)^2\\
&+\Big(\frac{3Y_{n}-4Y_{n-1}+Y_{n-2}}{2\triangle t}-\sum_{i=1}^{M}p_iF_i\big(X(n,:)\big)\Big)^2\Bigg\}
\end{align*}
and ODE models whose fitness values are too low will be eliminated in the process of genetic programming.
\subsection{Parameter Identification}
Once the ODE-structure is obtained, there are three groups of methods for solving the minimization problem (\ref{Jfp}): local, global and hybrid optimization methods (\cite{ashyraliyev2009systems}).
Orthogonal least square algorithm, as a global method, can analytically  determine parameters for linear-in parameter models. In this article we use this method to obtain the optimal model parameters. The idea of OLS algorithm is as follows:

   Mark $F$ and $P$ as
$$
F=
\left(
\begin{array}{ccc}
F_1(X(t_1))&\cdots&F_M(X(t_1))\\
\vdots&\ddots&\vdots\\
F_1(X(t_N))&\cdots&F_M(X(t_N))
\end{array}
\right),
P=
\left(
\begin{array}{c}
p_1\\
\vdots\\
p_M
\end{array}
\right).
$$
Then the parameter indentification equals solving vector  $P$, which meets $dY=FP$. Here $dY$ is the measured output vector, defined as Eq.(\ref{Y}); $F$ is the regression matrix, where $M$ is the number of regressors describing the basic unit of $f$  and $N$ is the length of vector $dY$.

As illustrated in \cite{madar2005genetic}, the OLS assumes that $F$ can be factorized as $$F=QR,$$
where $Q$ is an $N\times M$ orthogonal matrix and the columns of $Q$ are orthogonal satisfying $Q^{T}Q=D$, and $R$ is an $M\times M$ upper triangular matrix. Therefore,
$$Q^{T}dY=Q^{T}FP=Q^{T}QRP=DRP,$$
the OLS auxiliary parameter vector is $g=D^{-1}Q^{T}dY$, and the parameters in vector $P$
are readily computed from $$RP=g.$$

In practice, although some elements of $P$
exist, an overly small value of the element contributes little to the performance of the model. Therefore, we calculate the contribution of every function item  corresponding to $p_i$. Denote
\begin{equation}
\label{aa}
dY=FP+e,
\end{equation}
where $e$ is the error vector. After inserting $FP=QRP=Qg$ in to (\ref{aa}), it is easy to get
   $$(dY)^TdY=\sum_{i=1}^{M}g_i^2q_i^{T}q_i+e^Te,$$
   where $q_i$ is the column vector of $Q$, $g_i$ is the element of vector $g=D^{-1}O^{T}dY$.
   Define $Del_i=g_i^2q_i^{T}q_i/dY^TdY$. If $Del_i$ is less than the value 0.05 we predefined, we regard the corresponding $p_i$ as zero.
\section{Experimental results and prediction analysis}
The research data used in this study cover 120 days from January 21, 2016 to May 19, 2016, in Wuhan, China. The training set contains the former 100 days from
January 21, 2016, to April 29, 2016. And the data from April 30, 2016, to June 13, 2016, is the test set.
To validate the ODE models proposed in this paper, we compare the prediction results with the typical statistical model in the view of in-sample prediction and out-of-sample prediction.
\subsection{Statistical model}
Consider the $PM_{2.5}$ concentration from January 21, 2016, to April 29, 2016, as the training data.
By applying unit root test, the time series is  stationary; therefore, we use a typical $AR(p)$ model for the data. We apply the well-known Akaike information criterion(AIC) (\cite{akaike1998information}) and obtain the order $p=1$. Therefore, the $AR(1)$ model can describe the time series.
We perform out-of-sample  one-step-ahead prediction for April 30, 2016, and the real concentration of $PM_{2.5}$ is 47.1. The statistical model $AR(1)$  obtianed through EViews 8 is
\begin{equation}
\label{eq5}
 y_{k+1}=0.660439y_{k}+23.49820,
\end{equation}
 whose statistical results of out-of-sample  one-step-ahead prediction are shown in table \ref{tab1}.
\begin{table}[!htbp]
  \centering
    \begin{tabular}{lr}
    \toprule
    Prediction results & 54.7198 \\
    Root Mean Squared Error & 0.225989 \\
    Mean Absolute Error & 0.225989 \\
    Mean Abs. Percent Error & 0.475767 \\
    \bottomrule
    \end{tabular}%
      \caption{Out-of-sample  one-step-ahead prediction prediction of statistical model for April 30, 2016}
  \label{tab1}%
\end{table}%
\subsection{ODE models obtained by our data-driven method}
Meanwhile, we use data from January 21, 2016, to April
29, 2016, to train our ODE models and make a prediction for
 April 30, 2016. The experiment parameters are shown in table \ref{tab3}. Generation gap equals 0.8, which means individuals with the top 20\%  fitness value are selected as the parents of the next generation. When the number of generations reaches 20th, the evolution terminates.
\begin{table}[htbp]
  \centering
    \begin{tabular}{lc}
    \toprule
     Innitial population size & 30 \\
    Initial max tree depth & 5 \\
    Max Generation & 20 \\
    Crossover rate & 0.7 \\
    Generaton gap & 0.8 \\
    Mutation rate & 0.3 \\
    \bottomrule
    \end{tabular}%
      \caption{Experiment parameters for ODE-construction.}
  \label{tab3}%
\end{table}%

Because the ODE-construction method that we proposed in this paper is stochastic, each performing  maybe to get different ODEs. We perform the experiment 1,000 times and all the models and prediction results are as follows:
\begin{align}
ODE_1:& \frac{dy}{dt}=-0.152040, Pre=47.3480, Ape=0.005265393,\label{eq6}\\
ODE_2:&\frac{dy}{dt}=-0.113800, Pre= 47.3862, Ape=0.006076433,\label{eq7}\\
ODE_3:&\frac{dy}{dt}=5.537091*sin(t)-0.409219, Pre=50.3057, Ape=0.068061571,\notag\\
ODE_4:&\frac{dy}{dt}=0.091441*y*sin^3(t)-0.292838, Pre=48.0573, Ape=0.020324841,\notag\\
ODE_5:&\frac{dy}{dt}=0.000846*y^2*sin^3(t)-0.270239, Pre= 47.6034, Ape=0.010687898,\notag\\
ODE_6:&\frac{dy}{dt}=0.000006*y^3*sin^3(t)-0.265039, Pre=47.3608, Ape=0.005537155,\notag
\end{align}
\begin{figure}[!htbp]
\centering
 \includegraphics[width=0.6\textwidth]{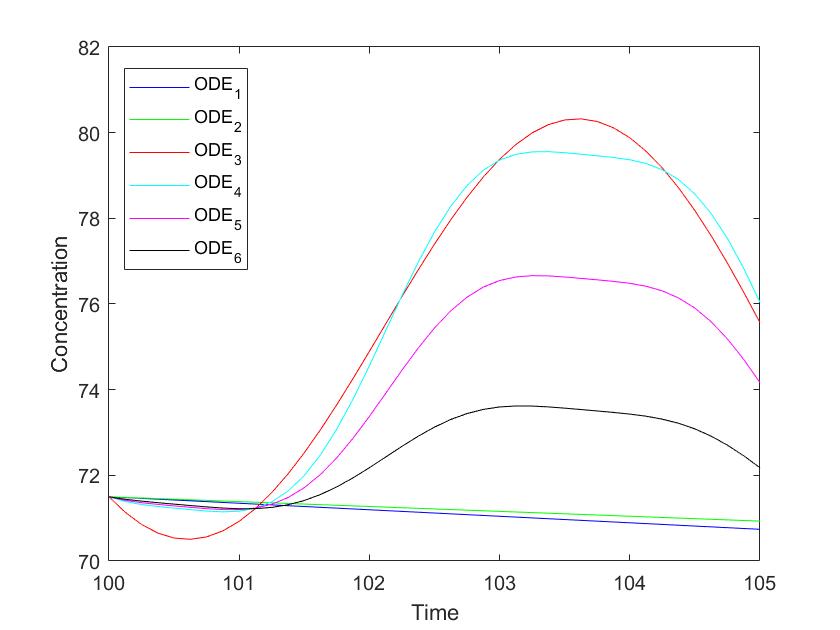}
\caption{Plots of ODE models for April
30, 2016 as shown the value y corresponding 101 of the x-axis.}
\label{fig5}
\end{figure}
where $Pre$ stands for the prediction results; $Ape$ is the absolute percent error, namely the absolute relative error. The specific prediction process is as follows: After we use 100 days of data from January 21, 2016, to April
29, 2016, to obtain the models above, we apply the concentration of $PM_{2.5}$ on April
29, 2016, as the initial value and predict the concentration of $PM_{2.5}$ on April
30, 2016, just as the predicted y value corresponding to time 101 of x-axis in Figure \ref{fig5}. Also,
we can see that (\ref{eq6}) and (\ref{eq7}) are essentially two linear polynomial models and their discrete forms are AR(1) models.
As each preforming generates different ODE models, we compute the expectation for the prediction results of one thousands times of performings and compare it with the real date as shown in table \ref{tab2}.
\begin{table}[htbp]
  \centering
    \begin{tabular}{lr}
    \toprule
    Expectation (Prediction results) & 47.88686 \\
    Root Mean Squared Error & 0.764535 \\
    Mean Absolute Error & 0.786863 \\
    Mean Abs. Percent Error & 0.017034 \\
    \bottomrule
    \end{tabular}%
  \caption{Out-of-sample  one-step-ahead prediction of ODE model for April 30, 2016}
  \label{tab2}%
\end{table}%

In this part, we list only the ODE models for predicting the concentration of $PM_{2.5}$ on April 30, 2016.
The ODE-models for predicting other days are listed in the supplementary materials.
By observing these models given in the supplementary materials, it can be seen that although genetic programming is a stochastic optimization method, the ODE models are different when performing programs each time. In this paper, we only select several simple models from this procedure for prediction. We will derive a systematic procedure to determine
the best model for prediction in the future.

\subsection{Prediction comparison between statistical model and the ODE models}
\begin{figure}[!htbp]
\centering
 \includegraphics[width=0.8\textwidth]{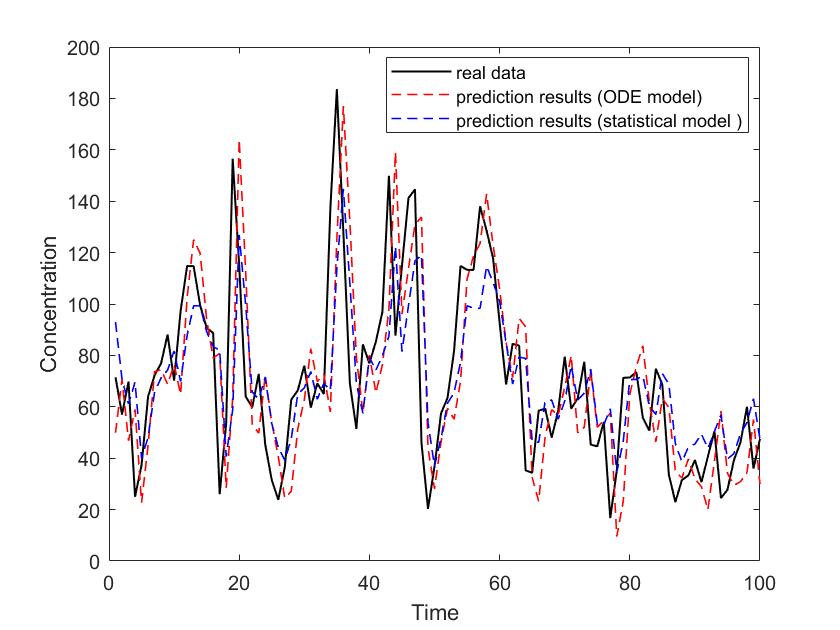}
\caption{In-Sample one-step-ahead for January 21, 2016 to April
29, 2016.}
\label{fig1}
\end{figure}
As seen from tables \ref{tab1} and \ref{tab2}, the ODE prediction models for April 30, 2016 are slight better than AR(1) model in the view of mean absolute percent error, but are worse in the view of root mean squared error and mean absolute error.

In the statistical field, in-sample and out-of-sample predictions are two points of view from which to measure the models for prediction. Therefore, below we will compare our ODE models with the traditional statistical model. Here, in-sample prediction is done to estimate the model with all the observations, and then we use the obtained model to predict some of the observations. For out-of-sample prediction, we divide the total observation into two parts. One part is to build the model and then to predict the other part of the data with the obtained model.
\begin{figure}
\centering
\includegraphics[width=0.6\textwidth]{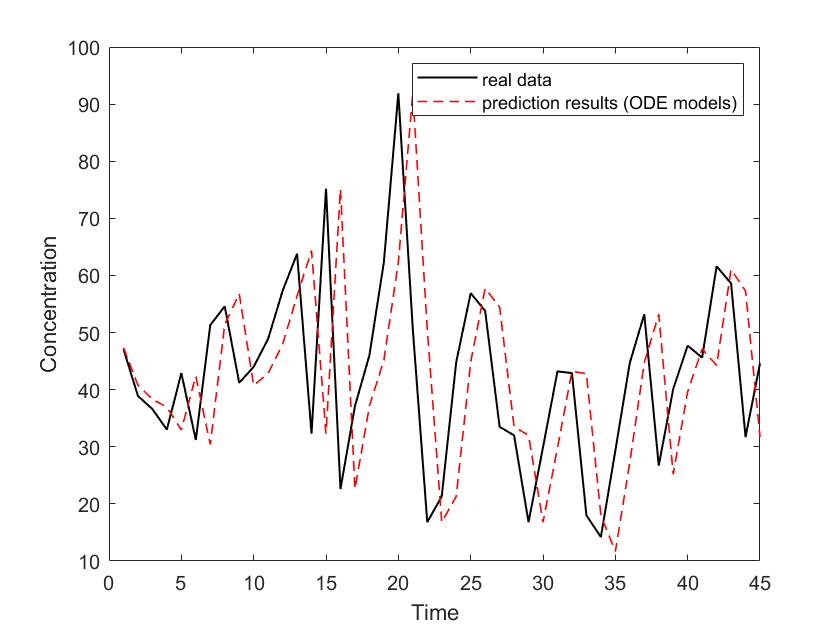}
\caption{Out-of-sample one-step-ahead  prediction from
 April 30, 2016 to June 13, 2016 by ODE-construction genetic program.}
\label{fig2}
\end{figure}

Figure \ref{fig1} shows the in-sample prediction results. Specifically, we use the data of 100 days from January 21, 2016, to April 29, 2016, to train models, and then we use the obtained model to make one-step ahead prediction for the same time period.
It is clear that there is a consistent trend between predicted data (represented by red lines and green lines) and real observations (presented by black lines).
In particular,  compared with the typical statistical model,  the prediction results obtained by the proposed ODE models are fairly good.

 Figure \ref{fig2} shows the out-of-sample one-step-ahead prediction results through the ODE models  proposed in this paper. Specifically, we use the data from January 21 to April 29, 2016, to train models and then use the obtained models to make one-step-ahead prediction from April 30 to June 13, 2016.
It is clear that there is a consistent trend between predicted data (represented by red lines) and real observations (represented by black lines).

As a result, the experiments show that the real-time ODE models are effective in approaching the real-time modeling and predicting tasks of series.
\section{Conclusion}
In this paper, ODE models are proposed to predict the daily $PM_{2.5}$ concentration. Tree-based genetic programming and least square method are employed to evolve the structure and model parameters of ODEs. The proposed method is based on observed real data and needs almost no meteorological assumptions; therefore, it can easily be applied to other problems. The experiment
results clearly illustrate that the ODE model can effectively predict the daily concentration of $PM_{2.5}$.

However, some issues need further discussion in our future work, as includes:

1) In practical application, although our ODE construction algorithm does not need meteorological knowledge or the specific mathematical formulation of the ODE model, some control parameters are predefined before making ODE construction, such as the function set, the operator set, the maximum tree depth, the mutation rate, the crossover rate and so on.  We will develop a rule to choose control parameters in future research.

2) The data-driven ODE model obtained in the current work is not unique, as the procedure involves a stochastic inputs. We will develop a systematic approach to determine the best model for the data.

3) Compared with ODE, a partial differential equation (PDE) involves a spatial dimension to describe the interplay between individuals, thus better describing the dynamic system in the spatial-temporal dimensions.
Therefore, a PDE model may better describe the transboundary pollution of $PM_{2.5}$. A PDE construction method should be developed in our future work.

\section*{Acknowledgments}
The authors would like to thank the editor and referees for their helpful comments which improve the paper. This work is supported by Scientific Research Project of Tianjin Municipal Education Commission (2017SK108).
\section*{References}

\end{document}